\documentclass[twocolumn]{aa}
\usepackage{graphicx,epsf,amssymb}
\def\be{\begin{equation}}
\def\ee{\end{equation}}
\def\bea{\begin{eqnarray}}
\def\eea{\end{eqnarray}}

\def\z{\zeta}
\def\p{\partial}

\begin{document}
\title{The Kelvin-Helmholtz Instability
    in an Expanding Universe and its Effect on Dark Matter}

\author{Karim A.~Malik\inst{1} \and David R.~Matravers\inst{2}}

\institute{ 
GRECO, Institut d'Astrophysique de Paris, C.N.R.S.,
98bis Boulevard Arago, \\
75014 Paris, France\\
\and
Institute of Cosmology and Gravitation,\\
University of Portsmouth, Portsmouth~PO1~2EG,
United Kingdom}

\date{astro-ph/0208272v3\ \ \hfill \ \ A\&A {\bf 406}, 37-41 (2003)}

\abstract{
We extend the Kelvin-Helmholtz instability to an expanding
background. We study the evolution of 
a non-viscous irrotational fluid and find that for wavelengths
much smaller than the Hubble scale small perturbations of
the fluid are unstable for wavenumbers larger than a critical
value. We then apply this result in the early universe,
treating cold dark matter as a 
classical fluid with vanishing background pressure.
}
%
%
%\pacs{95.30.Lz \ \ 95.35.+d \ \ \hfill \ \
%astro-ph/0208272v2}
%
%\maketitle
\titlerunning{The Kelvin-Helmholtz Instability in an 
Expanding Universe\ldots}
\authorrunning{Malik and Matravers}
\maketitle
%%%%%%%%%%%%%%%%%%%%%%%%%%%%%%%%%%%%%%%%%%%%%%%%%%%%%%%%%%%
\section{Introduction}
%%%%%%%%%%%%%%%%%%%%%%%%%%%%%%%%%%%%%%%%%%%%%%%%%%%%%%%%%%%

In the standard cosmological model quantum fluctuations are
magnified from microscopic to cosmological scales during a period
of exponential expansion (inflation) and source curvature
perturbations with a scale-invariant or nearly scale-invariant
spectrum (\cite{LLBook1,LLBook2}). After inflation these 
perturbations act as seeds for the large scale structure 
of the universe: first, beginning during radiation domination, 
non-baryonic dark matter ``falls'' into the potential 
wells formed during inflation 
and later, during matter domination but after decoupling, radiation 
and  baryonic matter follow.

In this picture radiation and baryonic matter 
can ``fall'' into the potential
wells only after decoupling since before that their Jeans length
is larger than the Hubble radius. After decoupling perturbations
whose wavelength is larger than the Jeans-length grow, but on
scales smaller than the Jeans-length they oscillate, leaving
imprints in the CMB that can be observed today. On very small
scales structure is wiped out through the tight coupling of the
baryons to the photons until decoupling (\cite{Silk}).
Unlike in baryonic matter and radiation, perturbations 
in the non-baryonic dark matter can start
growing as soon as their wavelength is smaller than the horizon,
since the Jeans scale is negligible for non-interacting dark matter.
The  non-baryonic dark matter can therefore cluster and 
``fall'' into the potential wells much earlier than the other matter.

In fluid dynamics the study of instabilities (\cite{chandra}) is
well established and has also been extended to the astrophysical
context (see \cite{ino}, and references therein). The
Kelvin-Helmholtz instability has been known for nearly 150 years
(\cite{chandra}). In its simplest form two large fluid volumes,
separated by an interface, move in opposite directions. Neglecting
the stabilizing effects of density and gravity gradients this
setup is unstable for all non-zero velocities and wavenumbers.

We study the initial stages in the evolution of small perturbations in
an irrotational, non-viscous fluid due to a Kelvin-Helmholtz
instability in an expanding universe. We show that even in such a
background these instabilities occur.
We then apply these results in the early universe. Two large
colliding regions of cold dark matter might serve as an example: the
Kelvin-Helmholtz instability can arise at the contact surfaces of the
two regions.
We model the dark matter as a fluid 
with vanishing background
pressure in concordance with the standard cold dark matter models
(\cite{LLBook1,LLBook2}). 
Depending on when this happens in the evolution of the
background the perturbations arising from the instability grow
either exponentially or linearly on scales smaller than a critical
wavenumber which we determine.
On the basis of these results we argue that it will be interesting
to test these perturbations in simulations because they could 
have relevance to the dark matter cusp and halo problems and to 
anisotropies in the CMB.

In the next section we give the governing equations 
for the fluid. In Sec.~\ref{modelsec} we specify the
model geometry and the general setup. 
After giving the background solution in Sec.~\ref{backsec},
we derive the perturbed equations in Sec.~\ref{pertsec}
and analyse the stability of the system. 
In Sec.~\ref{critwavesec} we then calculate the critical 
wavenumbers for which instability can occur. 
We apply our results to standard cold dark matter in the 
final section
%Sec.~\ref{discsec}
and discuss possible implications.

%%%%%%%%%%%%%%%%%%%%%%%%%%%%%%%%%%%%%%%%%%%%%%%%%%%%%%%%%%%
\section{Governing equations}
\label{govsec}
%%%%%%%%%%%%%%%%%%%%%%%%%%%%%%%%%%%%%%%%%%%%%%%%%%%%%%%%%%%

We are interested in the dynamics  of a fluid in the early universe on
scales much  smaller than the  horizon size, i.e.~$\lambda \ll  H$. We
can therefore use a Newtonian approximation on an expanding background
where only the  background expansion is described by  a flat FRW model
(\cite{LLBook1,LLBook2}).

We assume that the  fluid is non-viscous and
irrotational, its dynamics therefore governed by the Euler equation.
The fluid does not have to be  
the only or dominant matter component in the model of the 
universe, we merely require that it does not interact
with the other fluids such as radiation.
As outlined above we envisage this fluid
to describe standard cold  dark  matter, which has vanishing 
background pressure.
We shall nevertheless keep the discussion in this and the following
sections  applicable to the general case of a perfect fluid 
and restrict the analysis to
cold dark matter only when we consider a particular example 
in the final section.
%
%From Eq.~(\ref{dotrho}) we see that we can proceed with our 
%analysis without specifying 
%an equation of state for our fluid. Hence our results will hold for all 
%types of perfect fluid, including cdm.

The Euler equation in an expanding background is given by
(\cite{peebles})
\be
\label{Euler} 
\frac{\p}{\p t} u^i+H u^i+u^k\frac{\p}{a\p x^k} u^i
+\frac{1}{\rho}\frac{\p}{a\p x^i} P +\frac{\p}{a\p x^i}\phi=0 \,,
\ee
where $u^i$ is the three-velocity of the fluid, $P$ is the
pressure, $\rho$ is the energy density of the fluid,
$\phi$ is the gravitational potential, $a=a(t)$ is the scale
factor and $H\equiv\dot a / a$ is the Hubble rate
\footnote{
Notation: a dot denotes differentiation with respect to
coordinate time $t$, and $i,j=1,2,3$. Comoving length scales
$\lambda$ are related to the wavenumber by
$k\equiv2\pi/\lambda$.}.
The energy conservation equation in an expanding background
is given by
\be
\label{energy}
\frac{\p}{\p t} \rho +3H \rho
+ \frac{\p}{a\p x^k}\left(\rho u^k\right)=0 \,.
\ee
The condition for the fluid to be irrotational is,
\be
\label{irrot}
\frac{\p}{\p x^i} u^i =0 \,.
\ee
%

%%%%%%%%%%%%%%%%%%%%%%%%%%
\section{General formulation}
\label{modelsec}
%%%%%%%%%%%%%%%%%%%%%%%%%%

We can split the variables describing our problem into
a time-dependent background part and time- and
space-dependent first order perturbations,
\be
\label{split}
\psi=\psi(t)+\delta\psi(t,x^i)\,,
\ee
where $\psi\equiv\rho, \phi, P, u^i$.

\begin{figure}
\begin{center}
\includegraphics[width=80mm]{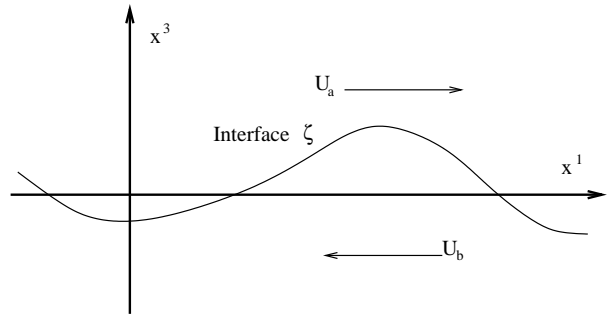} \\
\caption[interface]{\label{interface} The configuration: fluid
moving to the left with velocity $U_{\rm{b}}$ is separated from fluid
moving to the right with velocity $U_{\rm{a}}$
by the perturbed interface $\zeta$.}
\end{center}
\end{figure}

The geometry of our model is outlined in
Fig.~\ref{interface}: two large volumes
of fluid are separated by an interface at $x^3=0$ and move in
opposite directions with velocities $U_{\rm a}$ and $U_{\rm b}$
above and below the interface, respectively. We allow for a
perturbation of the fluid interface separating the two flows which
is denoted by $\zeta=\zeta(t, x^1, x^2)$.

The velocity can also be split according to Eq.~(\ref{split}),
$u^i = U^i+v^i$, where $U^i$ is the background part
and $v^i$ the perturbation.
Without loss of generality we can choose our coordinate system
such that the $x^1$-coordinate is aligned with the direction of
the background flow as in Fig.~\ref{interface}. The background
flow is therefore separated by the $x^3=0$ plane. The background
velocity $U^i$ has then only one non-zero component $U^1=U(t)$,
but for the velocity perturbation we still have $v^i=v^i(t,x^i)$.

We neglect gravitational back reaction on the perturbations and
therefore set $\delta\phi\equiv 0$, i.e.~the gravitational potential
$\phi$ only has a time-dependent zeroth order part.
%

%%%%%%%%%%%%%%%%%%%%%%%%%%%
%%%%%%%%%%%%%%%%%%%%%%%%%%

%%%%%%%%%%%%%%%%%%%%%%%%%%%%%%
\section{Background solutions}
\label{backsec}
%%%%%%%%%%%%%%%%%%%%%%%%%%%%%%

For the background described in the previous section
we get from Eqs.~(\ref{Euler}) and (\ref{energy})
the simple background Euler and energy conservation
equations
\bea
&&\dot U +HU =0 \label{dotU} \,, \\
&&\dot\rho+3H\rho=0 \label{dotrho} \,,
\eea
which have the solutions
\be
\label{backU} 
U=U_0 \left(\frac{a}{a_0}\right)^{-1} \,, \qquad
\rho=\rho_0 \left(\frac{a}{a_0}\right)^{-3} \,,
\ee
where $U_0$ and $\rho_0$ are constants of integration such that at
$t=t_0$ we have $U=U_0$ and $\rho=\rho_0$. 

The background expansion is described by a FRW model.
We therefore get a power law solution for the
scale factor (\cite{LLBook1,LLBook2})
\be
\label{scale} 
a=a_0\,\left(\frac{t}{t_0}\right)^\gamma \,,
\ee
where the constant of integration is chosen such that $a=a_0$ at
$t=t_0$. The exponent $\gamma$ characterizes the different epochs
of the early and late universe: $\gamma=1/2$ during radiation
domination and $\gamma=2/3$ during matter domination.
The evolution of the scale factor is governed by the
total background energy density. 
The solution (\ref{scale}) is therefore consistent with our model
in the matter dominated
regime and in the radiation dominated regime despite the fact that
the dark matter fluid always scales as $\rho \propto  a^{-3}$.
%$\rho \propto  a^{-3(1+w)}$.

%%%%%%%%%%%%%%%%%%%%%%%%%%%%%%
\section{Stability analysis}
\label{pertsec}
%%%%%%%%%%%%%%%%%%%%%%%%%%%%%%

In order to analyze the stability of our model
we first linearize the governing equations
and then look for growing mode solutions.

Linearizing around the background described in
Sec.~\ref{modelsec} we get
from Eq.~(\ref{Euler}) the linearized perturbed
Euler equation to first order
\be
\label{linEuler}
\frac{\p}{\p t} v^i+H v^i+U^1\frac{\p}{a\p x^1} v^i
+\frac{1}{\rho}\frac{\p}{a\p x^i} \delta P
%+\frac{\p}{a\p x^i}\phi
=0 \,,
\ee
and the energy conservation equation (\ref{energy})
becomes
\be
\label{lin_energy}
\dot{\delta\rho}+3H\delta\rho+U^1\frac{1}{a}\frac{\p}{\p x^1}\delta\rho
+\rho_0\frac{1}{a}\frac{\p}{\p x^k}v^k =0 \,.
\ee
The condition for the flow to be irrotational,
Eq.~(\ref{irrot}), becomes
\be
\label{cont}
\frac{\p}{\p x^i} v^i =0 \,.
\ee
Substituting Eq.~(\ref{cont}) into Eq.~(\ref{lin_energy}), we see
that the velocity perturbations decouple from the density
perturbations.
To analyse the Kelvin-Helmholtz instability it is therefore
sufficient to study Eqs.~(\ref{linEuler}) and (\ref{cont}).

The perturbation in the position of the interface,
given by $x^3=\z(t,x^1,x^2)$, is a first order
quantity and has to be included in our analysis.
The velocity of the fluid interface
in the 3-direction is given by
\be
\label{kinematic} v^3 =  a\left[\frac{\p\z}{\p t} +H\, \z
+u^1\frac{\p \z}{a\p x^1 } +u^2\frac{\p \z}{a\p x^2} \right] \,,
\ee
where
the RHS is simply the material derivative of the
physical fluid interface position. Linearizing around
the background described in Sec.~\ref{modelsec}
\be
\label{linkin}
v^3
=  a\left[\frac{\p\z}{\p t} +H\, \z
+U^1\frac{\p \z}{a\p x^1 } \right] \,.
\ee
We now make an ansatz for the first order quantities, extending
Chandrasekhar (\cite{chandra}),
\be
\label{ansatz} \delta Q(t,x^i) \propto e^{i\left(
k_1x^1+k_2x^2\right) +\omega(t)} \,, \ee
where $k_1$ and $k_2$ are the components of the wavevector and
$\omega$ is the \emph{time dependent} frequency.
With this ansatz growth of the perturbations and hence possible
instability will set in if $\omega(t) >0$.
Using the ansatz (\ref{ansatz}) the linearized Euler equations
(\ref{linEuler}) become
\bea 
\label{v1-v3}
\left(\dot\omega+H+iU^1\frac{k_1}{a}\right)v^{\alpha} &=&
-\frac{i}{\rho}\frac{k_{\alpha}}{a} \delta P \,, \nonumber \\
\left(\dot\omega+H+iU^1\frac{k_1}{a}\right)v^3 &=&
-\frac{1}{a\rho} \frac{\p}{\p x^3}\delta P \,, 
\eea
where $\alpha=1,2$, whereas the continuity equation (\ref{cont})
and the linearized kinematic condition Eq.~(\ref{linkin}) give,
respectively,
\bea
\label{cont_ans}
i\left(k_1v^1+k_2v^2\right)+\frac{\p}{\p
x^3}v^3 = 0 \,, \nonumber \\
\label{kin_ans}
v^3=\left(\dot\omega+H+iU^1\frac{k_1}{a}\right)a\z
\,.
\eea
The system of equations (\ref{v1-v3}) and (\ref{kin_ans}) can be
solved for the velocity perturbation $v^3$,
\be
\label{D2k2}
\left(\dot\omega+H+iU^1\frac{k_1}{a}\right)
\left(\frac{\p^2}{{\p x^3}^2}-{\bar k}^2\right) v^3=0 \,,
\ee
where  ${\bar k}^2=k_1^2+k_2^2$.
We see from Eq.~(\ref{D2k2}) that the $x^3$ dependent part of the
solution to $v^3$ has to be $\propto e^{\pm \bar{k} x^3}$.
The kinematic condition (\ref{kin_ans}) shows that
$v^3/\left(\dot\omega+H+iU^1\frac{k_1}{a}\right)$ has to be
continuous across the interface.
%We can therefore write down
The solution for $v^3$ is therefore given by
\bea \label{solution} v^3_{\rm a} &=&
A\left(\dot\omega+H+iU^1_{\rm a}\frac{k_1}{a}\right) e^{-\bar k
x^3} \qquad {\rm{for}} \; x^3 >0 \,, \nonumber \\ v^3_{\rm b} &=&
A\left(\dot\omega+H+iU^1_{\rm b}\frac{k_1}{a}\right) e^{\bar k
x^3}  \qquad {\rm{for}} \; x^3 < 0\,, \eea
where we have used the condition that the velocity has to stay
finite as $x^3\to \pm \infty$.

We now integrate Eq.~(\ref{D2k2}) across the interface from
$\zeta-\epsilon$ to $\zeta+\epsilon$ and then let $\epsilon \to 0$
to get
\bea \label{integ}
&&\left(\dot\omega+H+iU^1_{\rm{a}}\frac{k_1}{a}\right)
\frac{\p}{\p{x^3}}v^3_{\rm{a}} \nonumber \\
&\qquad&\qquad-\left(\dot\omega+H+iU^1_{\rm{b}}\frac{k_1}{a}\right)
\frac{\p}{\p{x^3}}v^3_{\rm{b}} =0 \,. 
\eea
Combining Eqs.~(\ref{solution}) and (\ref{integ}) finally gives an
equation for the time derivative of $\omega$,
\be
\label{doto}
\dot\omega=-H-\frac{1}{2}\frac{k_1}{a}
\Big[
i\left(U_{\rm a}+U_{\rm b}\right)
\pm \left|U_{\rm b}-U_{\rm a}\right| \Big] \,.
\ee
Since we are studying the growing or decaying behavior of the
solution we are only interested in the real part of $\omega$. We
can therefore neglect the complex term in the square brackets in
Eq.~(\ref{doto}) above, because it would only contribute an
oscillatory part to the solution. We also see that
Eq.~(\ref{doto}) allows two solutions depending on which sign we
choose in front of the $\left|U_{\rm b}-U_{\rm a}\right|$ term. We
only consider the ``$-$'' sign in the following, since it can
already be seen that because the Hubble parameter $H$ is always
positive the ``$+$'' choice will only lead to damped solutions.

Note that we recover the standard result of Chandrasekhar
(\cite{chandra}), i.e.~that all wavenumbers are exponentially
unstable, if we set $a=1$.

Equation (\ref{doto}) can now be integrated using the solution for
the background velocity Eq.~(\ref{backU}) and the scale factor
Eq.~(\ref{scale}). We have $\left|U_{\rm {b}}-U_{\rm {a}}\right|
=\bar U (t/t_0)^{-\gamma} $ where we have
defined $\bar U \equiv \left|U_{\rm {b0}}-U_{\rm
{a0}}\right|$.
We choose
as initial condition that there is no amplification
at the beginning, i.e.~$\omega=0$ at $t=t_0$.

We study the case of a radiation dominated universe,
i.e.~$\gamma = 1/2$, and a universe with a different
exponent $\gamma$ separately.

Integrating the real part of Eq.~(\ref{doto})
we get for the case $\gamma =1/2$
\be
\label{omrad}
\omega=\frac{1}{2}
\left(\frac{k_1 \bar U t_0}{a_0}-1\right)\ln\tilde t \,,
\ee
where we have defined $\tilde t\equiv t/t_0$.

In the case where $\gamma \neq 1/2$ we get, by integrating the
real part of Eq.~(\ref{doto}),
\be
\label{omgamma} \omega=-\gamma \ln\tilde t
+\frac{k_1}{A\left(2\gamma-1\right)} \left[1-{\tilde
t}^{1-2\gamma}\right]\,,
\ee
where we have introduced the quantity
\be
\label{defA}
A\equiv\frac{2a_0}{\bar U t_0} \,,
\ee
because it plays a crucial role in defining
critical wavenumbers in the following section.

%%%%%%%%%%%%%%%%%%%%%%%%%%%%%%%
\section{Critical wavenumber}
\label{critwavesec}
%%%%%%%%%%%%%%%%%%%%%%%%%%%%%%%

We can now analyze the solutions we have found in the previous
section and calculate the critical wavenumber beyond which the
perturbations are growing.

%%%%%%%%%%%%%%%%%%%%%%%%%%%%%%%%%%%%%%%%%%%%%%%%%%
\subsection{Radiation domination, $\gamma = 1/2$}
%%%%%%%%%%%%%%%%%%%%%%%%%%%%%%%%%%%%%%%%%%%%%%%%%%

In the radiation dominated regime, $\gamma =1/2$,
we find from Eq.~(\ref{omrad}) and our ansatz (\ref{ansatz})
that the flow is linearly unstable for all wavenumbers with
\be
\label{k1rad}
k_1 > k_{\rm crit,rad}\equiv\frac{A}{2} \,.
\ee
{}From Eqs.~(\ref{omrad}) and (\ref{k1rad}) we see that during
radiation domination perturbations with wavenumbers $k_1 > k_{\rm
crit,rad}$ only grow linearly, i.e,  $\delta Q \propto t$.

%%%%%%%%%%%%%%%%%%%%%%%%%%%%%%%
\subsection{Case $\gamma \neq 1/2$}
%%%%%%%%%%%%%%%%%%%%%%%%%%%%%%%

In the case of $\gamma \neq 1/2$
the behavior of the perturbations is more
complicated than in the radiation dominated case,
since here the critical wave number is time dependent.

The solution Eq.~(\ref{omgamma}) has two parts with
different behaviors:
the logarithmic part will always lead to a decrease in $\omega$,
whereas the second term behaves differently depending
on whether $\gamma$ is smaller or larger than 1/2.
In the case $\gamma <1/2$ the second term is always
increasing and will dominate after a finite time for
large enough wavenumbers.
For $\gamma >1/2$ the second term in Eq.~(\ref{omgamma}) will
first increase and then decrease but can lead to instability if it
dominates for a long enough period.
{}From Eq.~(\ref{omgamma}) the critical wavenumber
for which small perturbations begin to grow is
\be
\label{k1gamma}
k_1> k_{\rm crit,\gamma}\equiv
\gamma A
\frac{\left(2\gamma-1\right)\ln {\tilde t}}{1-{\tilde t}^{1-2\gamma}}\,.
\ee
Note that the limit $\tilde t \to 1$ of the RHS of
Eq.~(\ref{k1gamma}) above is
well defined,
$\lim_{\tilde t\to 1}\ln{\tilde t}/(1-{\tilde t}^{1-2\gamma})=1$.

We  will now discuss the two cases
$\gamma<1/2$ and $\gamma>1/2$ separately.

%%%%%%%%%%%%%%%%%%%%%%%%%%%%%%%%%%
\subsubsection{Case $\gamma<1/2$}
%%%%%%%%%%%%%%%%%%%%%%%%%%%%%%%%%%

The case $\gamma<1/2$ is particularly simple. Using the relation
(\cite{AS})
\be
\label{lninequ} 
\ln x \le n(x^{1/n}-1)\, \qquad {\rm for} \; x,n>0
\,,
\ee
we see that the time dependent part of Eq.~(\ref{k1gamma}) is
decreasing and we therefore get, as a conservative value for the
unstable wavenumber,
\be
\label{k1gamma<1/2} 
%k_1> 
k_{\rm crit,L}>
\gamma A \,.
\ee
%
%where the subscript L denotes the case $\gamma<1/2$.
For wave numbers satisfying this condition $\omega>0$ and the perturbations
grow exponentially with time as $\sim \exp({t^{1-2\gamma}})$ and so
the flow is exponentially unstable to small perturbations.

%%%%%%%%%%%%%%%%%%%%%%%%%%%%%%%%%%%%
\subsubsection{Case $\gamma > 1/2$}
%%%%%%%%%%%%%%%%%%%%%%%%%%%%%%%%%%%%

{}From Eq.~(\ref{k1gamma}) we see that for large $\tilde t $  the
time dependent term $\ln{\tilde t}/(1-{\tilde t}^{1-2\gamma})$
tends to $\ln \tilde t$. Hence, in this case,  the critical wave
number increases with time. In order to get a conservative
estimate of the critical wavenumber we evaluate
Eq.~(\ref{k1gamma}) at the time when a given perturbation is
largest.

The perturbations reach their maximum value at
\be
\label{tmax}
t_{\rm{max}}=t_0 \left(
\frac{k_1}{\gamma A}
\right)^\frac{1}{2\gamma-1} \,.
\ee
We can now evaluate Eq.~(\ref{k1gamma}) at
$t=t_{\rm{max}}$ and get an implicit equation
for the critical wavenumber for $\gamma>1/2$
\be
\label{k1gammacrit} 
\frac{k_{\rm crit,G}}{\gamma A}- \ln\left(
\frac{k_{\rm crit,G}}{\gamma A} \right)=1\,.
\ee
Solving Eq.~(\ref{k1gammacrit}) numerically shows that the error
in neglecting the logarithmic term is $\sim 0.1\%$.
We can therefore neglect it
and get as an estimate that the perturbations will grow
for wavenumbers
\be
\label{k1estim}
k_1 \gtrsim \gamma A \,.
\ee
When the perturbations reach their maximum
they have grown as
$\delta Q(t_{\rm{max}})=\delta Q(t_0) e^{\omega_{\rm{max}}}$,
where $\omega_{\rm{max}}=\omega(t_{\rm{max}})$. Note that we have
chosen $\omega(t_0)=0$. We therefore find that the perturbations
have grown at time $t_{\rm{max}}$ by a factor of
$e^{\omega_{\rm{max}}}$, where $\omega_{\rm{max}}$ is
given by
\be
\label{ommax}
\omega_{\rm{max}}=\frac{\gamma}{2\gamma-1}
\left[
\frac{k_1}{\gamma A}
-\ln\left(\frac{k_1}{\gamma A}\right)
-1\right] \,.
\ee
{}From Eq.~(\ref{lninequ}) we see that in Eq.~(\ref{ommax})
the square bracket is dominated by the first term
and hence the perturbations are amplified
by a factor of $\sim e^{k_1/A}$.

Our model covers the irrotational, non-viscous case. The
inclusion of viscosity is even more complicated: 
for small Reynolds numbers
it tends to dampen the instability, whereas for large Reynolds
numbers the development of turbulence becomes more likely.
Although the simplest CDM model does not include 
viscosity there are models,
like neutralino dark matter studied in (\cite{dominik}), which are
viscous. We note these instabilities will probably not be
detected in models based on collision-free Boltzmann distributions
and that the fluid model we use will break down at some small
scale dependent on the dark matter properties.

%%%%%%%%%%%%%%%%%%%%%%%%%%%%%%%%%%%%%
\section{Discussion and Conclusions}
\label{disksec}
%%%%%%%%%%%%%%%%%%%%%%%%%%%%%%%%%%%%%

In standard fluid dynamics the Kelvin-Helmholtz instability in the
incompressible case develops for all wavenumbers $k>0$. However in
an expanding universe we have shown that this is no longer the
case and that there are critical wave numbers which separate
stable and unstable domains. These wave numbers depend on the
expansion of the background and the  background velocity.  For
wave numbers larger than the critical wavenumber,
$k>k_{\rm{crit}}$, small perturbations will grow.

More specifically, for a scale factor exponent $\gamma > 1/2$
perturbations will grow for a wavenumber-dependent time. In the
case $\gamma=1/2$, i.e.~during radiation domination, perturbations
grow linearly for all times and for $\gamma <1/2$ they grow
exponentially for all times.

As an illustration we look at the case of cold dark matter 
at the beginning of matter domination, $\gamma=2/3$.
The physical setting could be two large volumes of cold dark matter 
colliding or some non-homogeneous velocity flows in the early 
universe.  In the first case the contact surface would be prone to the 
Kelvin-Helmholtz instability and in the second the instability would 
occur inside the flow.
From Eq.~(\ref{defA}) and (\ref{k1estim}) the
physical critical wavenumber at the time of matter and radiation
equality is given by
\be
\frac{k_{\rm {crit}}}{a_{\rm{eq}}} =\frac{2\gamma}{\bar U
t_0}\frac{a_0}{a_{\rm{eq}}} \,,
\ee
where $a_{\rm{eq}}$ is the scale factor at equality. We choose
$t_0$ to be the present age of the universe, $t_0=H_0^{-1}$, where
$H_0=100kms^{-1}Mpc^{-1} h$, ${a_0}/{a_{\rm{eq}}}=24000\Omega_0h^2$
(see (\cite{LLBook1,LLBook2})), with $\Omega_0\sim 1$ and the
characteristic velocity is $\bar U\sim 100kms^{-1}$  (see
\cite{anne,klypin}). With these numbers the critical
scale below which perturbations are amplified is
\be
\lambda_{\rm{crit}} \sim 1 kpc \,. 
\ee

Although detailed numerical analysis 
will be necessary in order to make definite quantitative
predictions of the effect of the instability on the density 
distribution, we can nevertheless speculate that it will lead 
to a redistribution of power from scales $\sim \lambda_{crit}$ 
to much smaller scales.

A significant feature of this example is that the instability
occurs at such a small scale. This is important because it is well
known that the cold dark matter model has problems at small
scales. It seems to predict higher densities and more small scale
structure than is observed. Modifications of the CDM have 
been tried, e.g. Warm Dark Matter and Self-Interacting
Dark Matter but neither of these solves all the problems, see
e.g.~(\cite{primack,tasitsiomi}) for an outline of the ideas and detailed
references.  
%As more observations have become available and as the
%numerical modelling of the small scale physics has become more
%detailed the fits have improved but there is still a residual
%problem.
%
While these discrepancies may be, at least
partially, due to numerical and observational resolution and
selection effects or the consequences of baryonic physics, there
still appears to be a residual problem.

The surprising element of the Kelvin-Helmholtz effect is that it
occurs even on expanding backgrounds and depending on the value of
$\gamma$ the instabilities can grow exponentially.  Our results
indicate that further analysis using numerical methods is needed
to explore the effect into the nonlinear regime.

%%%%%%%%%%%%%%%%%%%%%%%%%%%%
\begin{acknowledgements}
The authors would like to thank Anne Green, Andrew Liddle, 
Roy Maartens, Dominik Schwarz, David Wands and David Weinberg for useful
discussions.
KM is supported by a Marie Curie Fellowship
%of the European Community Programme
%%``\emph{Improving Human Research Potential and the Socio-economic
%%Knowledge Base}''
under the contract number \emph{HPMF-CT-2000-00981}.
\end{acknowledgements}
%%%%%%%%%%%%%%%%%%%%%%%%%%%%

%%%%%%%%%%%%%%%%%%%%%%%%%%%%
% Bibliography
%%%%%%%%%%%%%%%%%%%%%%%%%%%%
%

%%%%%%%%%%%%%%%%%%%%%
%
%
%%%%%%%%%%%%%%%%%%%%%%%%%%%%%%%%%%%%%%%%%%

\begin{thebibliography}{}
%
\bibitem[Abramowitz and Stegun, 1972]{AS}
M.~Abramowitz and I.~A.~Stegun,  {\it Handbook of Mathematical
 Functions}, 9th Edition, Dover Publications (1972).
%
\bibitem[Chandrasekhar, 1981]{chandra}
S.~Chandrasekhar, {\it Hydrodynamic and Hydromagnetic Stability},
Dover Publications (1981).
%
\bibitem[Green, 2002]{anne}
A.~M.~Green, 
%``Effect of halo modelling on WIMP exclusion
%limits,''
Phys.\ Rev.\ D {\bf 66} (2002) 083003
[arXiv:astro-ph/0207366].
%
\bibitem[Hofmann et al., 2001]{dominik}
S.~Hofmann, D.~J.~Schwarz and H.~St\"ocker,
%``Damping scales of neutralino cold dark matter,''
Phys.\ Rev.\ D {\bf 64} (2001) 083507
[arXiv:astro-ph/0104173].
%
\bibitem[Inogamov, 1999]{ino}
N.~A.~Inogamov,
%The Role of the Rayleigh-Taylor and Richtmyer-Meshkov
%Instabilities in Astrophysics: An Introduction
Astrophys.~Space Phys., Vol.~10, 1-335 (1999).
%
\bibitem[Klypin et al., 2001]{klypin}
A.~Klypin, Y.~Hoffman, A.~V.~Kravtsov and S.~Gottl\"{o}ber,
%''Constrained Simulations of the Real Universe: the Local
%Supercluster, '' 
[arXiv:astro-ph/0107104].
%
\bibitem[Kolb and Turner, 1990]{LLBook1}
%\bibitem{rocky}
E.~W.~Kolb and M.~S.~Turner, {\it The Early Universe},
{Addison-Wesley (1990) (Frontiers in physics,
69)}.
%
\bibitem[Liddle and Lyth, 2000]{LLBook2}
A.~R.~Liddle and D.~H.~Lyth, {\it Cosmological Inflation And
Large-Scale Structure,} Cambridge University Press (2000).
%
\bibitem[Peebles, 1980]{peebles}
P.~J.~E.~Peebles, {\it The Large Scale Structure of the
Universe}, Princeton Univiversity Press (1980).
%
\bibitem[Primack, 2002]{primack}
J.~R.~Primack, 
%Status of cold dark matter cosmology,
[arXiv:astro-ph/0205391].
%
\bibitem[Silk, 1968]{Silk}
J.~Silk,
%``Cosmic Black Body Radiation And Galaxy Formation,''
Astrophys.\ J.\  {\bf 151} (1968) 459.
%
\bibitem[Tasitsiomi, 2002]{tasitsiomi}
A.~Tasitsiomi,
%``The Cold Dark Matter crisis on galactic and subgalactic scales,''
[arXiv:astro-ph/0205464].




%%%%%%%%%%%%%%%%%%%%%
\end{thebibliography}
\end{document}